\documentstyle[epsfig]{aipproc}
\begin{document} 
\title{Investigating the Light Scalar Mesons} 
\author{D.Black$^a$, A. H. Fariborz$^b$, S. Moussa$^a$, S. Nasri$^a$,
and J. Schechter$^a$}
\maketitle
\vskip .1cm{\it$^a$ Department of Physics, Syracuse University, Syracuse,
New York 13244-1130, USA.}\vskip .2cm{\it $^b$ Department of
Mathematics/Science, State University of New York Institute of Technology,
Utica, New York 13504-3050, USA.}
%\lefthead{LEFT head}
%\righthead{RIGHT head}
%\maketitle
\begin{abstract}

We first briefly review a treatment of the scalars in meson meson
scattering based on a non-linear chiral Lagrangian, with
unitarity implemented by a "local" modification of the scalar
propagators. It is
shown that the main results are confirmed by a treatment in the SU(3)
linear sigma model in which unitarity is implemented "globally". Some
remarks are made on the speculative subject of the
scalars' quark structure.
\end{abstract}

\section{Introduction}

The "hydrogen atom" problem of meson spectroscopy is the study of the
pion in terms of its fundamental constituents. Typically, this difficult
problem is finessed by using an effective
Lagrangian treatment of the composite field which includes the important
feature of (almost)
spontaneously broken chiral symmetry. Then one explains the presumed next
highest mass meson-- the rho-- as a $q{\bar q}$ bound state and continues
up the spectrum. But nowadays there is increasing
support for the existence of the old "sigma" resonance which may be
lighter than the rho.
If this is true it certainly seems worthwhile to pause and examine
the issue in detail. It is also a difficult
problem because the sigma is in an energy range just above where one
expects chiral perturbation theory to be practical but well below where
asymptotic freedom permits a systematic perturbative QCD expansion.

    In this talk a recent paper\cite{bfmns}
on the subject will be discussed. Other work
is  referenced
in that paper and in other contributions \cite{h2001} to this conference.
First, a brief review of our previous results
based on the non linear
chiral Lagrangian will be given. Then we try to check the form of
these results by using the linear sigma
model. This model, while less general, provides the usual physical intuition
about the problem as it contains a scalar nonet linked to the
pseudoscalars.

\section{Brief review of our previous work using the nonlinear chiral
Lagrangian}

{\bf Pi pi scattering}\cite{pipi}. It was noted that the $I=J=0$ partial
wave
amplitude up to about 1 GeV could be simply explained as a sum 
of four pieces: i. the current algebra
"contact" term, ii. the $\rho$ exchange diagram iii. a non Breit Wigner
 $\sigma$ pole
diagram and exchange, iv. an $f_0(980)$ pole in the background produced
 by the other
three. This is illustrated in a step by step manner for the real part
$R^0_0$
in Figs. 1, 2 and 3.
%%%%%%%%%%%%%%%%%%%%%%%%%%%%%%%%%%%%%%%%%%%%%%%%%%%%%%%%%%%%%%%%%%%%%%%%%%%%%%%%%%%%%%%%%%%%%%%%%%%%%%%%%%%%%%%%%%%%%%%%%%%%%%%%%%%%%%%%%%%%%%%%%%%%%%%%%%%%%%%%%%%%%%%%%%%%%%%%%%%%%%%%%%%%%%%%%%%%%%%%%%%
\begin{figure}
\centering
\epsfig{file=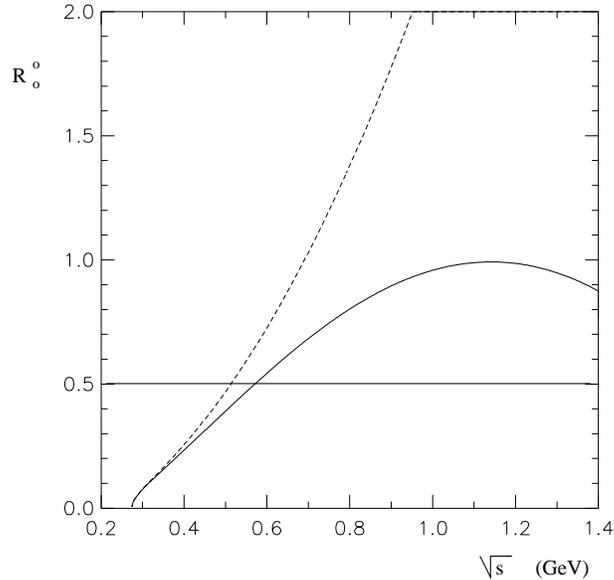,height=3in,angle=0}
\caption{The solid line which shows the current algebra $+ \rho$
result is much closer to the unitary bound of 0.5 than the dashed line
which shows the current algebra result alone.}
\end{figure}

\begin{figure}
\centering
\epsfig{file=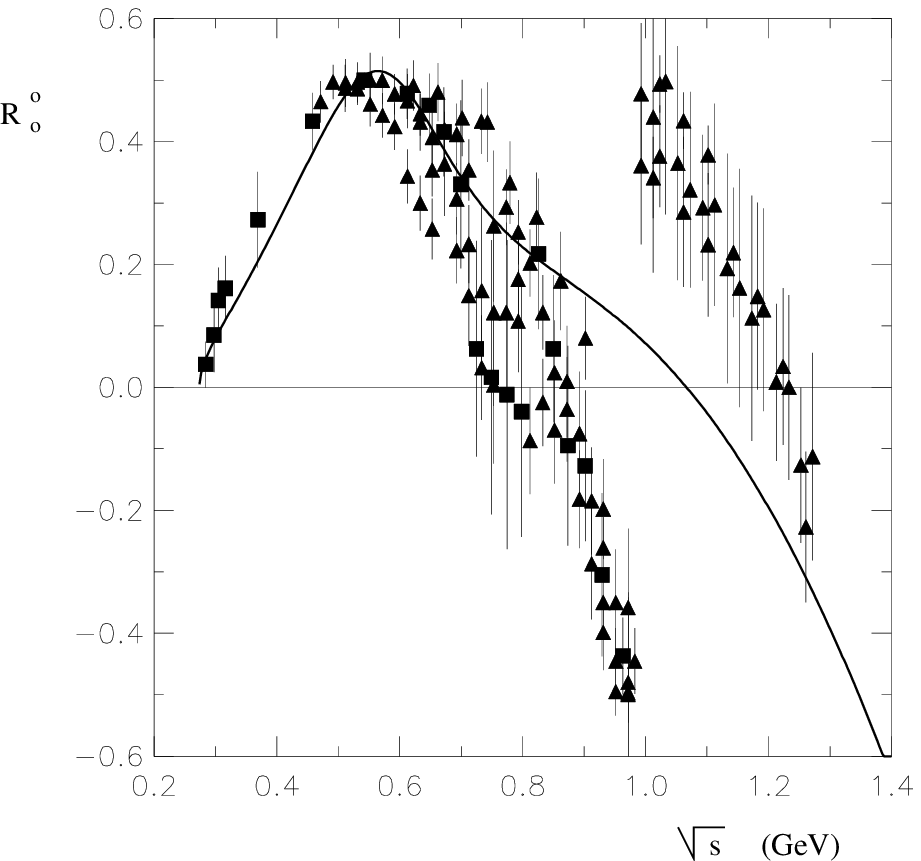,height=3in,angle=0}
\caption{The sum of current algebra $+ \rho + \sigma$ contributions
compared to data.}
\end{figure}

\begin{figure}
\centering
\epsfig{file=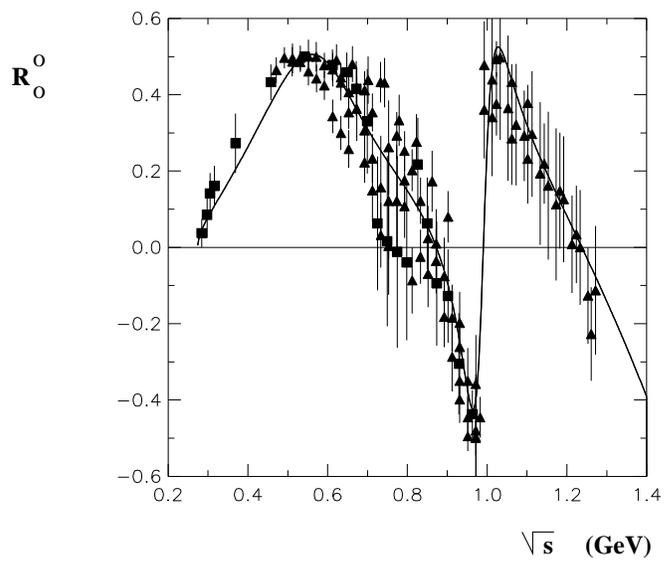,height=3in,angle=0}
\caption{The sum of current algebra $+ \rho + \sigma + f_0(980)$
contributions compared to data.}
\end{figure}

%%%%%%%%%%%%%%%%%%%%%%%%%%%%%%%%%%%%%%%%%%%%%%%%%%%%%%%%%%%%%%%%%%%%%%%%
We see in Fig. 1 that the ``current algebra" piece starts violating the
unitarity bound,  $  |R^0_0|  \leq 1/2 $ at about 0.5 GeV and
then runs away. However the inclusion of the  $\rho$
meson exchange diagrams turns the curve in the right direction and
improves, but does not completely cure, the unitarity violation.
These pieces, which do not involve any unknown parameters, give
encouragment to our hope that the cooperative interplay of various pieces
can explain the low energy scattering. In order to fix up
Fig. 1 we note that the real part of a resonance contribution
vanishes at the pole, is positive before the pole and {\it negative}
above the pole. Thus a scalar resonance with a pole roughly about 0.5 GeV
( above which $R^0_0$ in Fig. 1 needs a negative contribution to stay
below $1/2$) should do the job. The result of including such a $\sigma$
pole , with three parameters, is shown in Fig. 2. Now
note that the predicted $R^0_0(s)$ in Fig. 2 vanishes around 1 GeV.
Thus the phase $\delta$ at 1 GeV (assumed to keep rising) is
about $90^o$. Considering this as a background phase for the known
$f_0(980)$, the real part of the $f_0(980)$
contribution will get reversed in sign (Ramsauer--Townsend effect).
As Fig. 3 shows this is the missing piece needed to give a simple explanation 
of the $J=I=0$ $\pi\pi$ scattering up to about 1 GeV.

{\bf Pi K scattering}\cite{bfss}. In this case the low energy
amplitude is
taken to correspond to the sum of a current algebra contact diagram,
vector $\rho$ and $K^*$ exchange diagrams and scalar $\sigma(560)$,
$f_0(980)$ and $\kappa(900)$ exchange diagrams. The situation in the
interesting $I=1/2$ s-wave channel turns out to be very analogous to
the $I=0$ channel of s-wave $\pi\pi$ scattering. Now a non Breit Wigner
$\kappa$ is required to restore unitarity; it
plays the role of the $\sigma(560)$ in the $\pi\pi$ case. We found that
a satisfactory description of the 1-1.5 GeV s-wave region is also
obtained by including the well known $K_0^*(1430)$ scalar
resonance, which plays the role of the $f_0(980)$ in the $\pi\pi$
calculation. 

{\bf Putative light scalar nonet}\cite{putative}. The nine states
 associated with the $\sigma(560)$, $\kappa(900)$,
$f_0(980)$ and $a_0(980)$ are required in order to fit experiment 
in our model. What do their masses and coupling constants suggest
about their quark substructure? Clearly the mass ordering of the various 
states is inverted compared to the "ideal mixing"\cite{okubo}
scenario which approximately holds for most meson nonets. This means that 
a quark structure for the putative scalar nonet $N^b_a \sim q_a{\bar
q}^b$ is unlikely since the mass ordering just corresponds to counting the
number of heavier strange quarks. Then the degenerate $f_0(980)$ and $a_0(980)$
which must have the structure $N^1_1 \pm N^2_2$ would be lightest rather
than heaviest. However the inverted
ordering will agree with this counting if we assume that the scalar mesons
are schematically constructed as $N^b_a \sim T_a{\bar T}^b$ where $T_a
\sim \epsilon_{acd}{\bar q}^c{\bar q}^d$ is a "dual" quark (or
anti diquark). This interpretation is strengthened by consideration
\cite{putative} of the scalars' coupling constants to two 
pseudoscalars. That shows $\sigma \sim N^3_3 +$ "small", so it is
a predominantly non-strange particle in this picture. Furthermore
the states $N^1_1 \pm N^2_2$ now would each have two strange quarks
and would be expected to be heaviest. The four quark
picture was first suggested a long time ago\cite{jaffe} on dynamical
grounds.

{\bf Mechanism for next heavier scalar nonet}\cite{bfs}.
Of course, the success of the phenomenological quark model suggests
that there exists, in addition, a nonet of ``conventional"
p-wave $q{\bar q}$ scalars in
the 1+ GeV range. The experimental candidates for these states
are $a_0(1450)(I=1)$, $K_0^*(1430)(I=1/2)$ and for $I=0$,
$f_0(1370)$, $f_0(1500)$ and $f_0(1710)$. These are enough for a full 
nonet plus a glueball. However it is puzzling that the strange
$K_0^*(1430)$ isn't noticeably heavier than the non strange $a_0(1450)$
and that they are not lighter than the corresponding spin 2 states.
These and another puzzle may be solved in a natural way\cite{bfs}
if the heavier p-wave scalar nonet mixes with a lighter $qq{\bar q}
{\bar q}$ nonet of the type mentioned above. The mixing mechanism makes
essential use of the "bare" lighter nonet having an inverted mass ordering
while the heavier "bare" nonet has the normal ordering.
A rather rich structure involving
the light scalars seems to be emerging. At lower energies one may consider
as a first approximation,  "integrating out" the heavier nonet and retaining
just the lighter one. 

\section{The picture in a 3 flavor linear sigma model}

In \cite{bfmns} we employed the conventional chiral field $M=S+i\phi$,
where $S$ is a hermitian
matrix of nine scalars and $\phi$ is a hermitian matrix of nine pseudoscalars.
The Lagrangian density is,
\begin{equation}
{\cal L}=-\frac{1}{2}Tr(\partial_{\mu}M\partial_{\mu}M^{\dagger})
-V_0 +{\sum}A_a(M_{aa}+M_{aa}^{\dagger}).
\label{lagrangian}
\end{equation}
The three $A_a$'s are numbers proportional to the (current) quark
masses. $V_0$
may be considered to be an arbitrary function of the chiral $SU(3)\times
SU(3)$ invariants constructed from $M$ and $M^{\dagger}$. Note that most
consequences
at tree level follow just from chiral symmetry, irrespective of the form
of $V_0$.

{\bf Pi pi scattering amplitude}. The computed $I=J=0$ partial wave
amplitude at tree level has the form,
\begin{equation}
T^0_{0tree}(s)=cos^2\psi[\alpha(s)+\frac{\beta(s)}{m^2_{BARE}(\sigma)-s}]
+sin^2\psi[{\tilde \alpha}(s)+\frac{{\tilde \beta}(s)}
{m^2_{BARE}(\sigma^\prime)-s}],
\label{pipiamp}
\end{equation}
where $\alpha(s)$,$\beta(s)$ etc. are given in connection
with Eq (3.2) of \cite{bfmns}. $\psi$ is a mixing angle between
the two $I=0$ scalars, denoted as $\sigma$ and $\sigma^{\prime}$.
The subscript "BARE" on their masses means the value at tree level.
If $V_0$ is general, the three quantities $\psi$, $m_{BARE}(\sigma)$
and $m_{BARE}(\sigma^{\prime})$ may be chosen at will. However if
$V_0$ is taken to be renormalizable there is only one arbitrary parameter
(say $m_{BARE}(\sigma)$) in the theory when the input set (say $m_\pi,
m_K, m_\eta, m_{\eta^\prime}, F_\pi$) is fixed.

	It is instructive to first go back to the widely treated two
flavor case. This corresponds to choosing $\psi=0$ in (\ref{pipiamp}).
Then $m_{BARE}(\sigma)$ is the only unknown parameter. Near threshold,
if $m_{BARE}(\sigma)$ is not too low, the amplitude is the "current
algebra" result which agrees fairly well with experiment. It is a small
quantity which emerges from an almost complete cancellation of
 the pole and non pole terms in (\ref{pipiamp}). One
would like
to keep this result and utilize the effect of the sigma at higher energies.
Since there is a true pole in (\ref{pipiamp}) it seems reasonable to
regulate this in the usual way by adding a term $-im\Gamma$ in the pole
denominator. However, as Achasov and Shestakov \cite{as} pointed out,
this regulation completely destroys the good current algebra result.
They instead adopt the K matrix approach (whereas the usual solution is to
adopt the non linear model instead since the derivative coupling of the
$\sigma$ there suppresses the pole contribution near threshold).
 In this way the tree amplitude
is not only regularized but made exactly unitary. One calculates the amplitude 
in terms of its tree value as:
\begin{equation}
T=\frac{T_{tree}}{1-iT_{tree}} .
\label{kmatrix}
\end{equation}
When $T_{tree}$ is small, $T \approx T_{tree}$ so the behavior near
threshold will now not be spoiled. At the other extreme, when $T_{tree}$
gets very large $T \rightarrow i$.
     
Note that the pole position, $z$ of the unitarized $T$ will typically
correspond to  mass and width (via $z=m^2-im\Gamma$) which differ from
$m_{BARE}$ and the starting perturbative width. Which one should be
chosen? Since $T$ in (\ref{kmatrix}) evidently has the
structure of a "bubble sum" in field theory it seems reasonable to
regard the K matrix unitarization as an approximation to including
the "radiative corrections". Then, as in usual field theory, the pole
found is interpreted
as giving the physical mass and width while the values of $m_{BARE}$
and $\Gamma_{BARE}$ would have no special significance. For the
two flavor model we verified the result of \cite{as} that a choice of
$m_{BARE}(\sigma)$ around 0.8 to 1.0 GeV would result in a physical
$m(\sigma)$ around 0.45 GeV and fit the first bump in Fig. 3. The physical
mass is not very sensitive to the exact choice of bare mass and also the
physical width is very greatly reduced. The predicted mass
in this model is  a bit less than the one we found in the non linear model
reviewed in the previous section, but this is readily understandable as
being due \cite{hss2} to the neglect of vector mesons in the present
model.  

{\bf Three flavor linear model amplitude}. The procedure was simply to use
the full two pole
tree amplitude (\ref{pipiamp}) in the unitarization formula
(\ref{kmatrix}). We
were not able to fit the entire
$T^0_0$ amplitude shown in Fig. 3 up to about 1.2 GeV in the
renormalizable model (which contains only the single unknown parameter  
$m_{BARE}(\sigma)$). However it is easy to find a fit in the chiral
model with general $V_0$, in which we were able to choose the three
parameters $m_{BARE}(\sigma)= 0.847$ GeV, $m_{BARE}(\sigma^\prime)=
1.300$ GeV and $\psi= 48.6^o$. The physical isoscalar masses (after
unitarization) turned out to be 0.457 GeV and 0.993 Gev associated with 
respective widths 0.632 GeV and 0.05 GeV. Again these represent large
shifts from the bare values. For illustrative purposes the unitarized
amplitude is reasonably approximated as
\begin{equation}
T^0_0 \approx const. + \frac{0.167+0.210i}{s-(0.209-0.289i)}
+\frac{0.053+0.005i}{s-(0.986 -0.051i)}.
\label{polefit}
\end{equation}
Neither the first ($\sigma$) pole nor 
the second ($f_0(980)$) pole is precisely of Breit Wigner type.
However the $f_0(980)$  pole approximates a Breit Wigner
except for an overall minus sign, which corresponds to the 
well known "flipping" of this resonance.

We similarly studied the $I=0, J=1/2$ scattering amplitude to find
the properties of the $\kappa$ resonance in the linear model. The bare 
mass of the $\kappa$ is fixed once the input parameters are given.
To allow us to vary this quantity (in a range where the $\pi \pi$
scattering is not much affected) we chose the alternative input set
$(m_\pi, m_K, m_{\eta^\prime}, F_\pi,F_K)$ and varied $F_K$. In this case,
because the $K_0^*(1430)$ is not included in the model we can only fit the data
up to about 1 GeV. It was found that the best fit corresponded to the 
bare $\kappa$ mass about 1.3 GeV. After unitarization the physical kappa
mass turned out to be about 0.800 GeV and this didn't change much as the
bare value was varied from 0.9 to 1.3 GeV. Unitarization also
substantially narrowed the physical kappa width. Furthermore, as for the
case of the $\sigma$ the $\kappa$ pole is not of Breit Wigner type. 
An analogous calculation was carried out to study the properties of the 
$a_0(980)$ as observed in $\pi \eta$ scattering. A summary shown
in Table 1 compares
the physical widths obtained in this linear model with those
obtained in the non linear model. In the cases of the $f_0(980)$
and the $a_0(980)$ the entries were taken from the Particle Data Group
\cite{rpp}, with which the non linear model calculations agree.

\begin{table}
\begin{tabular}{lcccc}
 &  $\sigma$ & $f_0$ & $\kappa$  & $a_0$ \\ \tableline
Present Model& & & & \\
mass (MeV), width (MeV) & 457, 632 & 993, 51 & 800, 260-610 & 890-1010,
110-240 \\
Comparison & & & & \\
mass (MeV), width (MeV)& 560, 370 & 980$\pm 10$, 40-100 &
900, 275 & 985,  50-100 \\
\end{tabular}
\caption{Predicted ``physical'' masses and widths in MeV of the nonet
of scalar mesons contrasted with suitable (as discussed in the text)
comparison values.}
\label{summary_table}
\end{table}
   
    Clearly, the complex pole positions and nature of the poles (non
Breit Wigner for $\sigma$ and $\kappa$ and "Ramsauer Townsend"
for $f_0(980)$) of the scalar nonet in the linear sigma model
are similar to those obtained previously (putative scalar nonet) using
a non linear chiral Lagrangian with a different "local" regulation. This
statement makes heavy use of the unitarization of the three flavor linear
model; otherwise the $f_0(980)$ and $\kappa$ might be considered
too high and wide to belong to a light scalar nonet. In particular,
the $\kappa$ clearly cannot be identified with the $K_0^*(1430)$ in
this unitarized linear sigma model. 

{\bf Speculation on scalar quark structure} (Section V of \cite{bfmns}).
At an intuitive level one might expect the scalar nonet, being the "chiral
partner" of the light pseudoscalar nonet, to have a quark- anti quark
structure. It was stressed \cite{putative} however that in the
more general non linear
Lagrangian approach (e.g. \cite{ccwz}) the scalar and pseudoscalar
transformation
properties are decoupled. Only the flavor SU(3) transformation
property, not the chiral one, of the scalars is fixed in the effective
non-linear
Lagrangian treatment. Features, mentioned above, like isoscalar mixing
angle and mass ordering suggest in fact the $q q {\bar q} {\bar q}$
structure for the light scalars as an initial approximation.

     How might this kind of scenario play out in the linear model
where the chiral properties of the scalars and the pseudoscalars
are clearly linked?  Even there, the quark substructure implied by the 
$SU(3) \times SU(3)$ transformation properties of the chiral matrix $M$ in
(\ref{lagrangian})
is not unique \cite{bfmns} (However the $U(1)_A$ transformations do
distinguish between $q{\bar q}$ and $qq{\bar q}{\bar q}$). There are 
three different "four quark" structures with the same transformation
properties. Physically, they correspond to making the chiral mesons
as a)  meson meson "molecule" b) spin 0 diquark -spin 0 anti diquark
and c) spin 1 diquark - spin 1 anti diquark. Actually these three
are not linearly independent. Thus the molecule \cite{iw}
and diquark- anti diquark \cite{jaffe} pictures are not clearly
distinguished at the effective Lagrangian level. Presumably, large changes
in the properties of the scalars due to unitarization in the effective
theory must be counted as "four quark" admixtures at the underlying level.

    In detail,
the schematic structure for the matrix $M(x) = S + i \phi $ realizing a $q
\bar q$
composite in terms of quark fields $q_{aA}(x)$ can be written
\begin{equation}
 M_a^{(1)b} = {\left( q_{bA} \right)}^\dagger \gamma_4 \frac{1 +
\gamma_5}{2} q_{aA}, \label{M1} \end{equation} where $a$ and $A$ are
respectively flavor and color indices. For the "molecule" model a) the
schematic quark structure  with the same $SU(3)_L \times SU(3)_R$
transformation property is,
\begin{equation} M_a^{(2)b} = \epsilon_{acd}
\epsilon^{bef} {\left( M^{(1) \dagger} \right)}_e^c {\left( M^{(1)
\dagger}
 \right)}_f^d.
 \label{M2} 
\end{equation}
In the spin 0 diquark - spin 0 anti diquark case the same transformation
property is realized with, 
 \begin{equation}
 M_g^{(3)f} = {\left(L^{gA}\right)}^\dagger R^{fA},
\end{equation}
 where
 \begin{eqnarray} L^{gE} = \epsilon^{gab}
\epsilon^{EAB}q_{aA}^T C^{-1} \frac{1 + \gamma_5}{2} q_{bB}, \nonumber \\
R^{gE} = \epsilon^{gab} \epsilon^{EAB}q_{aA}^T C^{-1} \frac{1 -
\gamma_5}{2} q_{bB},
 \end{eqnarray}
in which $C$ is the charge conjugation matrix of the Dirac theory.
Finally the spin 1 diquark - spin 1 anti diquark case c) has the
schematic structure,
\begin{equation}
 M_g^{(4) f} = {\left( L^{g}_{\mu \nu,AB}\right)}^\dagger R^{f}_{\mu
\nu,AB}, \end{equation}
 where
 \begin{eqnarray} L_{\mu \nu,AB}^g = L_{\mu
\nu,BA}^g = \epsilon^{gab} q^T_{aA} C^{-1} \sigma_{\mu \nu}
 \frac{1 + \gamma_5}{2} q_{bB}, \nonumber \\ R_{\mu \nu,AB}^g = R_{\mu
\nu,BA}^g = \epsilon^{gab} q^T_{aA} C^{-1} \sigma_{\mu \nu}
 \frac{1 - \gamma_5}{2} q_{bB}, \end{eqnarray}
and $\sigma_{\mu \nu} = \frac{1}{2i} \left[ \gamma_\mu, \gamma_\nu
\right] $.

 Now, as discussed before, the realistic situation is likely to contain
substantial mixing between scalar $q{\bar q}$ and $qq{\bar q}{\bar q}$
nonets. To explore this we formulated a linear sigma model containing both
a $q{\bar q}$ chiral matrix $M = S + i\phi$ and a $qq{\bar q}{\bar q}$
chiral matrix
$M^\prime = S^{\prime} + i{\phi}^{\prime}$. This is a very complicated
system
 so we started with a "toy
model" in which all current quark masses are neglected and only a minimum
number of non derivative terms are included. In addition to two minimal
kinetic
terms as in (\ref{lagrangian}), we took the simplified potential,
\begin{equation} V_0 = -c_2 {\rm Tr} \left( M M^\dagger \right) + c_4 {\rm
Tr} \left( M M^\dagger M
 M^\dagger \right) + d_2 {\rm Tr} \left( M^\prime M^{\prime \dagger}
\right) +
 e {\rm Tr} \left( M M^{\prime \dagger} + M^\prime M^\dagger \right).
\label{mixingpot} \end{equation}
Here $c_2$, $c_4$ and $d_2$ are positive real constants.  The $M$ matrix
field is chosen to have a wrong
 sign mass term so that there will be spontaneous breakdown of chiral
symmetry.
 A pseudoscalar octet will thus be massless. The mixing is controlled
by the parameter $e$. It is amusing to note that there will then be an
induced $qq{\bar q}{\bar q}$ condensate $\left<S^{\prime}\right>$
in addition to the usual $q{\bar q}$ condensate $ \left<S\right> $.

 We found that it is easy to obtain a
situation where the the next highest state above the
predominantly $q{\bar q}$ Nambu Goldstone
pseudoscalar octet is a predominantly $qq{\bar
q}{\bar q}$ scalar octet. Still heavier is the predominantly $qq{\bar
q}{\bar q}$ pseudoscalar while heaviest of all is the $q{\bar q}$ scalar
octet. Of course, SU(3) symmetry breaking and unitarization would be
expected to modify this picture. It seems very interesting to further
pursue a model of this type. There is evidently a possibility of learning
a lot about non perturbative QCD from the light scalar system.

 We would like to thank
Francesco Sannino and Masayasu Harada for fruitful collaboration. One of
us (J.S.) would like to thank the organizers for arranging a stimulating
and enjoyable conference. The work has been supported in part by the US
DOE under contract DE-FG-02-85ER40231.
 

\begin{references}

\bibitem{bfmns}D. Black, A. H. Fariborz, S. Moussa, S. Nasri and
J. Schechter, Phys. Rev.{\bf D64}, 014031(2001). 
 
\bibitem{h2001}See the write-ups of N. Achasov, C. Gobel, S. Ishida,
M. Ishida, T. Kunihiro, J. L. Lucio-Martinez, G. Moreno, W. Ochs,
J. Pelaez, Yu. Surovtsev, K. Teshihiko, T. Teshima, S. F. Tuan
and E. van Beveren.

\bibitem{pipi}M. Harada, F. Sannino and J. Schechter, Phys. Rev.
{\bf D54},1991(1996).

\bibitem{bfss}D. Black, A. H. Fariborz, F. Sannino and J. Schechter,
Phys. Rev. {\bf D58}, 054012 (1998).
 
\bibitem{putative}D. Black, A. H. Fariborz, F. Sannino and J. Schechter,
Phys. Rev. {\bf D59}, 074026 (1999).

\bibitem{okubo}S. Okubo, Phys. Lett. {\bf 5}, 165 (1963).

\bibitem{jaffe} R. Jaffe, Phys. Rev. {\bf D15}, 267 (1977).

\bibitem{bfs}D. Black, A. H. Fariborz and J. Schechter, Phys. Rev.
{\bf D61}, 074001 (2000).

\bibitem{as}N. N. Achasov and G. N. Shestakov, Phys. Rev. {\bf D 49},
5779 (1994).

\bibitem{hss2}M. Harada, F. Sannino and J. Schechter, Phys. Rev. Lett.,
{\bf 78}, 1603 (1997).

\bibitem{rpp}Review of Particle Physics,
Euro. Phys. J. {\bf C3} (1999).

\bibitem{ccwz}C. Callan, S. Coleman, J. Wess and B. Zumino, Phys. Rev.
{\bf 177}, 2247 (1969).

\bibitem{iw}N. Isgur and J. Weinstein, Phys. Rev. {\bf D27}, 588 (1983).

\end{references}
\end{document}